\documentclass{sf2a-conf2018}
\usepackage{graphicx}
\usepackage{hyperref}
\usepackage[]{natbib}  
\usepackage{epstopdf}

\def\BibTeX{{\rm B\kern-.05em{\sc i\kern-.025em b}\kern-.08em
    T\kern-.1667em\lower.7ex\hbox{E}\kern-.125emX}}
\bibpunct{(}{)}{;}{a}{}{,}  

\usepackage{amsmath,bm}

\newcommand{\Ro}{\mathrm{Ro}}
\newcommand{\rmv}{\mathrm{v}}
\newcommand{\sun}{\odot}

\begin{document}

\TitreGlobal{SF2A 2018}


\title{Penetration of Rotating Convection} 
\runningtitle{Penetration of Rotating Convection}
\author{K.~C. Augustson}\address{AIM, CEA, CNRS, Universit\'{e} Paris-Saclay, Universit\'{e} Paris Diderot, Sarbonne Paris Cit\'{e}, F-91191 Gif-sur-Yvette Cedex, France}
\author{S. Mathis$^1$}

\setcounter{page}{237}


\maketitle


\begin{abstract}
  A simplified model for stellar and planetary convection is derived for the magnitude of the rms velocity, degree of
  superadiabaticity, and characteristic length scale with Rossby number as well as with thermal and viscous
  diffusivities. Integrating the convection model into a linearization of the dynamics in the transition region between
  convectively unstable and stably-stratified region yields a Rossby number, diffusivity, and pressure scale height
  dependent convective penetration depth into the stable region. This may have important consequences for mixing along
  the evolution of rotating stars.
\end{abstract}



\section{Introduction}\label{sec:intro}

The secular impacts of rotation and magnetic fields on stellar and planetary evolution are of keen interest within the
astrophysical community \citep[e.g.,][]{maeder09,mathis13a}. As expounded upon in \citet{stevenson79}, a surprisingly
effective approach to including rotation in MLT is to hypothesize a convection model where the convective length-scale,
degree of superadiabaticity, and velocity are governed by the linear mode that maximizes the convective heat flux. This
model of rotating convection has its origins in the principle of maximum heat transport proposed by \citet{malkus54}.
In that principle, an upper limit for a boundary condition dependent turbulent heat flux is established that depends
upon the smallest Rayleigh unstable convective eddy. The size of this eddy is determined with a variational technique
that is similar to that developed in \citet{chandrasekhar61} for the determination of the Rayleigh number, which is the
ratio of the buoyancy force to the viscous force multiplied by the ratio of the thermal to viscous diffusion
timescales. This technique then permits the independent computation of the rms values of the fluctuating temperature and
velocity amplitudes. Numerical simulations have lent some credence to this simple model \citep{kapyla05,barker14}. In
particular, those simulations indicate that the low Rossby number scaling regime established in \citet{stevenson79}
appears to hold up well for three decades in Rossby number ($\Ro$ is the ratio of inertial and Coriolis forces) and for
about one decade in Nusselt number ($\mathrm{Nu}$ is the ratio of the convective and conductive fluxes). What remains to
be shown is how such a model of convection can impact the mixing for intermediate Rossby numbers while including
diffusion and the depth of convective penetration.

\section{General Framework}

The heuristic model will be considered to be local such that the length scales of the flow are much smaller than either
density or pressure scale heights. This is equivalent to ignoring the global dynamics and assuming that the convection
can be approximated as local at each radius and colatitude in a star or planet.  As such, one may consider the dynamics
to be Boussinesq. In other words, the model consists of an infinte Cartesian plane of a nearly incompressible fluid with
a small thermal expansion coefficient $\alpha_T=-\partial\ln{\rho}/\partial T|_P$ that is confined between two
impenetrable plates differing in temperature by $\Delta T$ and separated by a distance $\ell_0$. As seen in many papers
regarding Boussinesq dynamics \citep[e.g.,][]{chandrasekhar61}, the linearized Boussinesq equations can be reduced to a
single third-order in time and eighth-order in space equation for the vertical velocity.  The difference here and in the
work of \citet{stevenson79} is that the state that the system is being linearized about is nonlinearly saturated,
meaning that the potential temperature gradient is given by the Malkus-Howard convection theory
\citep[e.g.,][]{malkus54,howard63}.  Together, these equations provide a dispersion relationship on which the convection
model can be constructed.  The details of how this model can be constructed are given in \citet{augustson18}.  The
parametric quantities needed to see how this model can be leveraged to give estimates of the rotational and diffusive
influence are

\vspace{-0.25truein}
\begin{center}
  \begin{align}
     z^3 = \frac{k^2}{k_z^2}, \quad
    O = q \sqrt{\frac{3}{2}}\frac{\cos{\theta}}{5\pi\Ro}= q O_0, \quad
    K = q \frac{\kappa k_z^2}{N_0} = q K_0,\quad
    V = q \frac{\nu k_z^2}{N_0} = q V_0, \label{eqn:equivalencies}
  \end{align}
\end{center}

\noindent where $k$ is the magnitude of the wavevector characterizing the mode that maximizes the heat flux, $k_z$ is
its vertical component, and $\theta$ is the latitude. Note that the variation of the superadiabaticity for this system
is given by $\epsilon = H_P \beta/T$, meaning that $N^2 = g \alpha_T T \epsilon/H_P$, where $H_P$ is the pressure scale
height and $N_0$ is the buoyancy flux of the nondiffusive and nonrotating system provided by the Malkus-Howard
convection theory, $\beta$ is the potential temperature gradient, $\kappa$ is the thermal diffusion coefficient, $\nu$
is the viscous diffusion coefficient, and $q = N_0/N$. The convective Rossby number is
$\Ro= \rmv_0 /(2\Omega_0 \ell_0)$, where $\Omega_0$ is the constant angular velocity of the system and the
characteristic velocity $\rmv_0$ is easily derived from the nonrotating and nondiffusive case as
$\rmv_0 = \frac{s_0}{k_0} = \sqrt{6} N_0\ell_0/(5\pi)$. Two relevant equations are the dispersion relationship linking
$\sigma$ to $q$ and $z$, and the heat flux $F$ to be maximized with respect to $z$

\vspace{-0.25truein}
\begin{center}
  \begin{align}
    &\left(\sigma \!+\! K_0 q z^3\right)\!\! \left(\! z^3\!\left(\sigma\!+\! V_0 q z^3\right)^2 \!\!+\! 4 O_0^2q^2\!\right)\!-\! \left(z^3\!-\! 1\right)\!\!\left(\sigma\!+\! V_0 q z^3\right)\!=\!0, \label{eqn:fullcharq}\\
    &\frac{F}{F_0} = \frac{1}{q^3} \left[\frac{\sigma^3}{z^3}+V_0 q\sigma^2\right].\label{eqn:heatfluxq}
  \end{align}
\end{center}

To assess the scaling of the superadiabaticity, the velocity, and the horizontal wavevector, a further assumption must
be made in which the maximum heat flux is invariant to any parameters, namely that $\max{\left[F\right]}=F_0$ so the
heat flux is equal to the maximum value $F_0$ obtained in the Malkus-Howard turbulence model for the nonrotating case.
Therefore, building this convection model consists of three steps: deriving a dispersion relationship that links
$\sigma$ to $q$ and $z$, maximizing the heat flux with respect to $z$, and assuming an invariant maximum heat flux that
then closes this three variable system.

\section{Convection Model}

In the case of planetary and stellar interiors, the viscous damping timescale is generally longer than the convective
overturning timescale (e.g., $V_0\ll N_0$).  Thus, the maximized heat flux invariance is much simpler to treat.
In particular, the flux invariance condition under this assumption is then

\vspace{-0.25truein}
\begin{center}
  \begin{align}
    \frac{\max{\left[F\right]}}{F_0} &=\left.\frac{\sigma^3}{q^3z^3}+\frac{V_0\sigma^2}{q^2}\right|_{\mathrm{max}}\approx
    \left.\frac{\sigma^3}{q^3z^3}\right|_{\mathrm{max}} =1 \implies \sigma=q z + \mathcal{O}(V_0/N_0).\label{eqn:maxndhf}
  \end{align}
\end{center}

One primary assumption of this convection model is that the magnitude of the velocity is defined as the ratio of the
maximizing growth rate and wavevector. With the above approximation, the velocity amplitude can be defined
generally. The velocity relative to the nondiffusive and nonrotating case scales as

\vspace{-0.25truein}
\begin{center}
  \begin{align}
    \frac{\rmv}{\rmv_0} &= \left(\frac{5}{2}\right)^{\frac{1}{6}}\frac{\sigma}{q z^{3/2}} = \left(\frac{5}{2}\right)^{\frac{1}{6}} z^{-\frac{1}{2}}.\label{eqn:vsteve}
  \end{align}
\end{center}

To find the scaling of the heat flux maximizing wavevector $k = z^{3/2}$ and the superadiabaticity
$\epsilon/\epsilon_0=q^{-2}$, one may find the implicit wavevector derivative of the growth rate $\sigma$ from Equation
\ref{eqn:fullcharq} and equate it to the derivative of the heat flux $\partial F/\partial z = \sigma/z$, which neglects
the heat flux arising from the viscous effects.  Using the heat-flux invariance, e.g. letting $\sigma = qz$, the
constraining dispersion relationship (Equation \ref{eqn:fullcharq}) can be manipulated to solve for $q$ as a function of
$z$.  Substituting this solution into the equation resulting from the flux maximization yields an equation solely for
the wavevector $z$:

\vspace{-0.25truein}
\begin{center}
  \begin{align}
    &z^3\! \left(V_0 z^2\!+\! 1\right)^2\! \left[3V_0 K_0 z^4\!\left(2 z^3\!-\! 3\right)+ z^2\!\left(V_0\!+\!
      K_0\right)\!\left(4z^3\!-\! 7\right)\!+\! 2 z^3\!-\! 5\right]\!\nonumber\\
    &\qquad\qquad-\!\frac{6 \cos^2{\!\theta}}{25 \pi^2 \Ro^2}\!\! \left[K_0\left(3V_0z^5\!+\! z^3\!+\! 2\right)\!+\!
      V_0\left(5z^3\!-\! 2\right)\!+\! 3z\right]=0,\label{eqn:zeqndiff}
  \end{align}
\end{center}

\noindent whereas the superadiabaticity is found to be

\vspace{-0.25truein}
\begin{center}
  \begin{align}
    &\frac{\epsilon}{\epsilon_0} \!=\! \left(\frac{2}{5}\right)^{\frac{2}{3}}\frac{\left(1\!+\!K_0 z^2\right)\! \left(25 \pi^2 \Ro^2
      z^5\!\left(1\!+\! V_0 z^2\right)^2\!+\! 6 \cos^2{\theta}\right)}{25\pi^2 \Ro^2 \left(z^3\!-\!
      1\right)\left(1\!+\! V_0 z^2\right)}. \label{eqn:xeqnfd}
  \end{align}
\end{center}

\section{Convective Penetration}\label{sec:penetration}

The \citet{zahn91} model of convective penetration is built upon a linearization of the thermodynamics with respect to
the vertical displacement, which permits the equation of motion to be integrated in depth from the point where the
convective flux vanishes to the point where the velocity vanishes. This yields an estimate of the depth of penetration
$L_P$ of a fluid element that depends upon the boundary value of the convective velocity. Following \citet{zahn91} as
closely as possible, one may consider the system at the pole so that the direct effects of the local Coriolis
acceleration $2\Omega_0\sin{\theta}\rmv_x$ may be neglected. Instead, the Coriolis effect implicitly influences the
penetration depth by modifying the upper boundary value of the velocity. From Equation 3.9 of \citet{zahn91}, the
penetration depth scales as

\vspace{-0.25truein}
\begin{center}
  \begin{align}
    &\frac{L_P}{H_P} \!=\! \left[\frac{2}{3}\frac{\left(1-f\right) f \rmv_z^{3}}{g \alpha_T \kappa\chi_P\nabla_{\mathrm{ad}}}\right]^{\frac{1}{2}},
  \end{align}
\end{center}

\noindent where $\rmv_z$ is the boundary value of the velocity given by the convection model derived above,
$\nabla_{\mathrm{ad}}$ is the adiabatic temperature gradient and $\chi_P=\partial \ln{\kappa}/\partial \ln{P}|_{S}$ is
the adiabatic logarithmic derivative of the radiative conductivity with respect to pressure. It is assumed that only
downward penetrating flows are effective at carrying enthalpy. This asymmetry between upflows and downflows is
parameterized through the filling factor $f$. Note that the adiabatic temperature gradient
$\nabla_{\mathrm{ad}} = dT/dz|_{\mathrm{ad}} + \epsilon$.  However, a basic assumption of the model is that the
superadiabaticity $\epsilon$ does not grow large enough to modify the background temperature gradient in a steady
state. Thus, the ratio of the penetration depth with rotation and diffusion to the nonrotating inviscid value for
convective penetration into a stable layer either above or below a convection zone therefore scales as

\vspace{-0.25truein}
\begin{center}
  \begin{align}
    &\frac{L_P}{L_{P,0}} \!=\! \left(\frac{\rmv}{\rmv_0}\right)^{3/2} = \left(\frac{5}{2}\right)^{\frac{1}{4}}z^{-\frac{3}{4}}.
  \end{align}
\end{center}

\begin{figure}[t!]
  \begin{center}
    \includegraphics[width=0.9\textwidth]{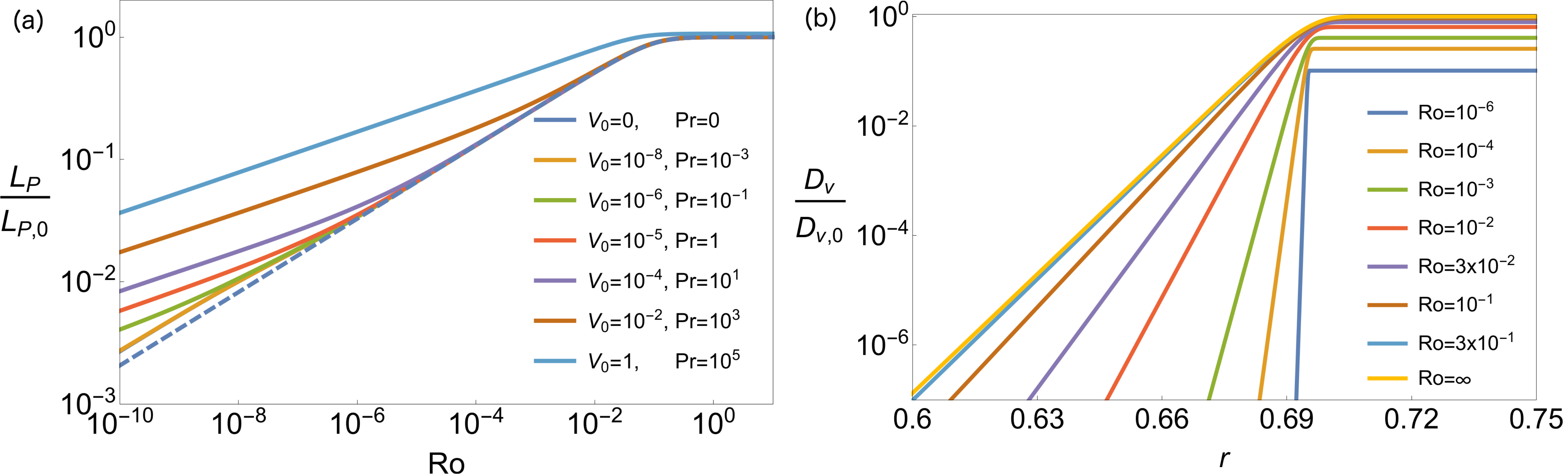} 
    \caption{Rossby and Prandtl number dependencies of the convective penetration depth $L_P$ at the pole
      ($\theta=0$). (a) Scaling of $L_P$ with viscosity $V_0$ at a fixed thermal diffusivity $K_0=10^{-5}$.  (b) The
      radial dependence of the vertical mixing length diffusion coefficient for a solar-like model for the inviscid
      convection model, showing the dual effects of decreased diffusion with decreasing Rossby number and the increasing
      lower radial limit of the diffusion coefficient due to the decreasing depth of penetration.}\label{fig:lp_scaling}
  \end{center}
\end{figure}

As seen in the previous section, the velocity amplitude of the mode that maximizes the heat flux decreases with lower
diffusivities and lower Rossby numbers. Therefore, the penetration depth necessarily must decrease when the Rossby
number is decreased. This behavior follows intuitively given that the reduced vertical momentum of the flows implies
that the temperature perturbations are also reduced. Thus, due to the decreased buoyant thermal equilibration time and
the reduced inertia of the flow the penetration depth must decrease. In contrast, the velocity and the horizontal scale
of the flow increase with greater diffusivities in order to offset the reduced temperature perturbations in the case of
a larger thermal conductivity.  In the case of a larger viscosity, the horizontal scale of the velocity field is
increased, whereas, for a fixed thermal conductivity, the thermal perturbations are of a smaller scale. Thus, to
maintain the heat flux, the amplitude of the velocity must increase in order to compensate for the reduced correlations
between the two fields. The scaling behaviors of the penetration depth are illustrated as a function of diffusivities
and Rossby number in Figure \ref{fig:lp_scaling}(a).

In the 3D f-plane simulations of rotating convection described in \citet{brummell02}, it is found that the penetration
depth into a stable layer below a convective region scales as $L_P \propto \Ro^{0.15}$, due primarily to a reduction in
the flow amplitude.  In a similar suite of f-plane simulations examined in \citet{pal07}, it is found that there is a
decrease in the penetration depth with increasing rotation rate that scales as $L_P \propto \Ro^{0.2}$ at the pole and
to $L_P \propto \Ro^{0.4}$ at mid-latitude.  The depth of convective penetration as assessed in those numerical
simulations appears to be roughly consistent with the heuristic model derived above, where
$L_P/L_{P,0} \propto \Ro^{3/10}$, which follows from $\rmv/\rmv_0 \propto \Ro^{1/5}$ in the nondiffusive and low Rossby
number limit of the convection model.

\section{A Diffusive Approach}

A diffusive parameterization of mixing processes has been extensively examined in many stellar settings. One such model
has been established through an extreme-value statistical analysis of 3D penetrative convection simulations
\citep{pratt17a}, permitting the construction of a model for a turbulent diffusion based upon the Gumbel distribution
\citep{pratt17b}.  Using the above extension of the \citet{zahn91} model, one can estimate both the penetration depth
and the level of turbulent diffusion as a function of the Rossby number and diffusivities of the convection model. Doing
so yields the following description of the radial dependence of the diffusion coefficient

\vspace{-0.25truein}
\begin{center}
  \begin{align}
  	&D_\rmv\left(r\right) =\left(\frac{5}{2}\right)^{\frac{1}{6}}\!\frac{\alpha H_{P} \rmv_c}{3\sqrt{z}} \left\{1\!-\!\exp{\!\left[-\exp{\left(\!\left(r-r_c\right)/\lambda L_P\!+\!\mu/\lambda\right)}\right]} \right\},\label{eqn:pdiff}
  \end{align}
\end{center}

\noindent where $\rmv_c$ and $r_c$ are the velocity and radius at the base of the convection zone and where $\mu$ and
$\lambda$ are the empirically determined parameters from \citet{pratt17b}.  An illustrative example of the scaling
behavior of $D_\rmv$ for a solar-like star where the transition region begins around $r\approx 0.7 R_{\sun}$ is shown in
Figure \ref{fig:lp_scaling}(b). The radial structure of the diffusion coefficient follows from the scaling of the
velocity, namely the diffusion will globally decrease with decreasing Rossby number.  The depth of penetration is
perhaps most notable, in that its strong rotational dependence can lead to severe restrictions on the region in which
the diffusion acts. This potentially has strong implications for mixing in rotating stars \citep[e.g.,][]{jorgensen18}.

\section{Summary}\label{sec:final}

A simple model of rotating convection originating with \citet{stevenson79} has been extended to include thermal and
viscous diffusion for any convective Rossby number.  Moreover, a systematic means of developing such models has been
developed for an arbitrary dispersion relationship.  An explicit expression is given for the scaling of the horizontal
wavenumber in terms of the Rossby number and diffusion coefficients (Equation \ref{eqn:zeqndiff}), from which a similar
scaling of the velocity and superadiabaticity is derived (Equations \ref{eqn:vsteve} and \ref{eqn:xeqnfd}). Utilizing
the linearized model of \citet{zahn91}, this rotating convection model is employed to assess the scaling of the depth of
convective penetration with Rossby number and diffusivities. The turbulent diffusivity arising from that penetrating
convection is then estimated utilizing the statistical model found in 3D simulations \citep{pratt17b,pratt17a}.

\begin{acknowledgements}{The authors acknowledge support from the ERC SPIRE 647383 grant and PLATO CNES grant at CEA/DAp-AIM.}\end{acknowledgements}

\bibliographystyle{aa}
\bibliography{augustson_convection}

\end{document}